\documentclass[aps, prl, preprintnumbers,amsmath,amssymb,latexsym,array,enumerate,letter,twocolumn,superscriptaddress]{revtex4-2}

\usepackage{amssymb}
\usepackage{amsmath}
\usepackage{epsfig}
\usepackage{hyperref}
\usepackage{xcolor}
\usepackage{slashed}
\usepackage{dsfont,bbm}
\usepackage{mathrsfs}

\usepackage{multirow, graphicx, subfigure, placeins, float}

\graphicspath{{./figs/}}

\begin{document}

\preprint{USTC-ICTS/PCFT-26-51}
\title{Three-Loop Five-Point CK-Dual Amplitudes and UV Structure \\ in ${\cal N}=4$ SYM and ${\cal N}=8$ SUGRA}
\author{Zhiming Cai}
\email{caizhiming@itp.ac.cn}
\affiliation{Institute of Theoretical Physics, Chinese Academy of Sciences, Beijing 100190, China}
\affiliation{School of Physical Sciences, University of Chinese Academy of Sciences, Beijing 100049, China}
\author{Zeyu Li}
\email{2506398088@pku.edu.cn}
\affiliation{Center for High Energy Physics, Peking University, Beijing 100871, China}
\author{Gang Yang}
\email{yangg@itp.ac.cn}
\affiliation{Institute of Theoretical Physics, Chinese Academy of Sciences, Beijing 100190, China}
\affiliation{School of Physical Sciences, University of Chinese Academy of Sciences, Beijing 100049, China}
\affiliation{School of Fundamental Physics and Mathematical Sciences, Hangzhou Institute for Advanced Study, UCAS, Hangzhou 310024, China}
\affiliation{Peng Huanwu Center for Fundamental Theory, Hefei, Anhui 230026, China}
\author{Guorui Zhu}
\email{zhuguorui@itp.ac.cn}
\affiliation{Institute of Theoretical Physics, Chinese Academy of Sciences, Beijing 100190, China}
\affiliation{School of Physical Sciences, University of Chinese Academy of Sciences, Beijing 100049, China}

\begin{abstract}
We construct the complete full-color three-loop five-point integrand of
$\mathcal N=4$ super-Yang--Mills theory in a representation that manifestly
satisfies color--kinematics duality. Its double copy gives the corresponding
$\mathcal N=8$ supergravity integrand. 
For four-dimensional external states, we evaluate the ultraviolet poles of
both amplitudes in the critical dimension, $D_c=6$. 
Extending the external-state dependence to $D$ dimensions is subtle.
We consider a candidate replacement of the four-dimensional prefactors by expressions
built from $D$-dimensional tree amplitudes.
It reproduces the open-string prediction for the SYM pole with generic $D$-dimensional external states, 
whereas the corresponding gravity expression differs from the string-inspired one by an evanescent term.
\end{abstract}

\maketitle

\section{Introduction}

Color--kinematics (CK) duality reorganizes gauge-theory amplitudes so that
kinematic numerators obey the same Jacobi relations as color factors
\cite{Bern:2008qj}. Replacing the color factors by a second set of dual numerators then
produces gravity amplitudes, realizing perturbative gravity as a double copy of
gauge theory \cite{Bern:2010ue, Bern:2010yg}. 
Although established at tree level \cite{BjerrumBohr:2009rd, Stieberger:2009hq,Feng:2010my}, CK duality remains conjectural at loop level and has been demonstrated only in explicit examples; see \cite{Bern:2019prr,Bern:2022wqg,Adamo:2022dcm} for reviews.

In $\mathcal N=4$ super-Yang--Mills (SYM) theory, the CK-dual 
four-point integrands are known through four loops \cite{Bern:2012uf}, and 
explicit full-color five-point amplitudes have been available through two loops
\cite{Carrasco:2011mn,Mafra:2015mja}. 
Moving simultaneously to three loops and
five points is qualitatively harder: nonplanar topologies proliferate, graph
symmetries overlap strongly with the Jacobi relations, and CK-dual numerators 
must satisfy a more extensive spanning set of generalized-unitarity cuts. 
The five-point amplitude also carries physical information absent at four points,
including a nontrivial dependence on the dimensionality of the external states 
and the nonlinear structure of the UV counterterm.

Here we present, to our knowledge, the first explicit full-color three-loop
five-point CK-dual integrand of $\mathcal N=4$ SYM and, by double copy, the
corresponding $\mathcal N=8$ supergravity (SUGRA) integrand. 
A three-loop construction was announced in \cite{Carrasco:2011mn}
but no explicit result was presented.
We perform two independent constructions using $S_5$ and $S_3 \times S_2$ symmetries, 
which yield the same amplitude after integrand reduction.
The CK-dual numerators, at most quadratic in loop momenta, give a compact representation that retains the complete nonplanar information.

We use these integrands to determine the five-point UV poles at $D_c=6$, the
critical dimension consistent with the three-loop four-point case \cite{Bern:2007hh}. 
The SYM pole contains no double-trace term and determines the five-field matrix element of the corresponding counterterm.
Replacing the four-dimensional external-state prefactors by expressions built from $D$-dimensional tree amplitudes,
we find that the SYM pole reproduces the open-string prediction for generic $D$-dimensional external states.
For gravity, however, the corresponding lifted UV pole differs from the string-inspired result by an evanescent term. 
This implies that the four-dimensional representation ceases to determine the full integrands of higher-dimensional theories,
and an interpretation of the mismatch in terms of tensorial projection onto four-dimensional external kinematics is discussed.

\section{CK-dual integrand construction}

\begin{figure}[t]
	\centerline{\includegraphics[height=1.6cm]{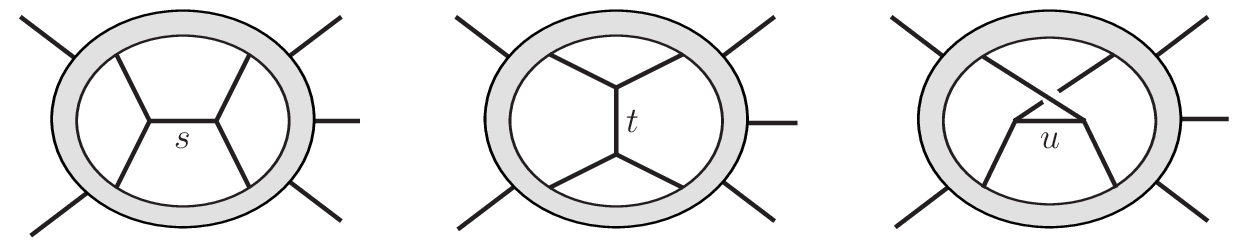} }
	\caption{Color-kinematics duality at loop level: the $s$-, $t$- and $u$-channel graphs are related by Jacobi identities.}
	\label{fig:loopBCJ}
\end{figure}

Our construction proceeds in two steps (see e.g.,~\cite{Bern:2012uf, Boels:2012ew} for the general strategy): (I) we build a CK-dual ansatz by expressing all numerators through a small set of master numerators via dual Jacobi relations; (II) we solve the ansatz by imposing diagrammatic symmetries and generalized-unitarity cuts.

\begin{figure*}[t]
\centering
\includegraphics[width= \textwidth]{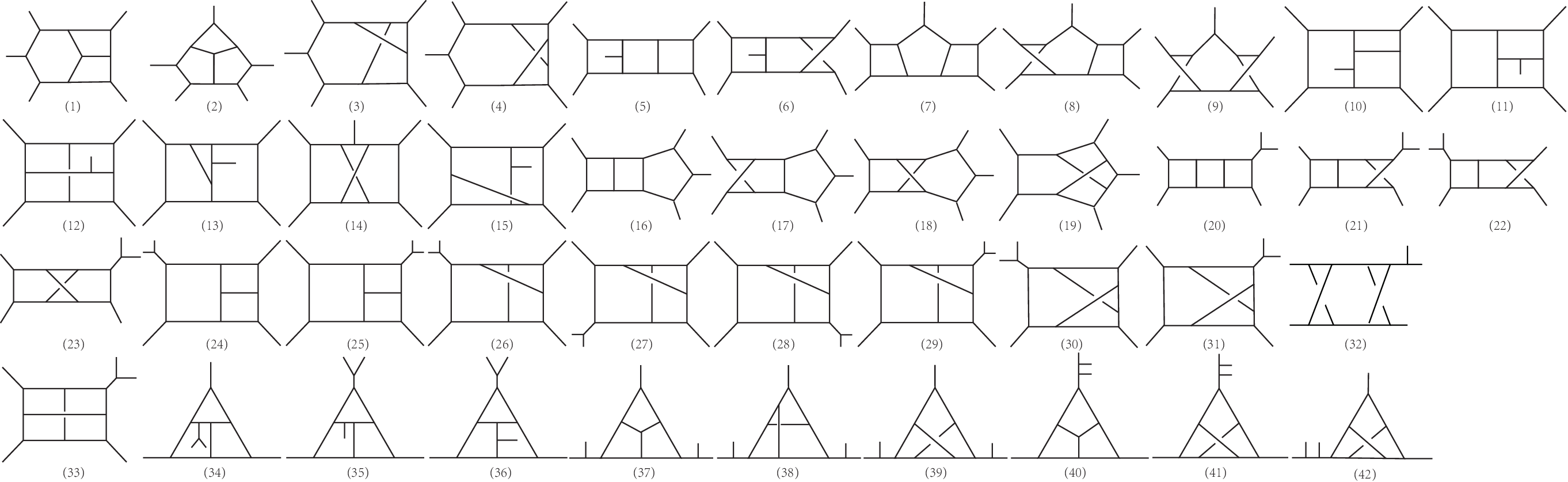}
\caption{All 42 non-isomorphic trivalent topologies for the five-point three-loop amplitude. The first two are chosen as masters.}
\label{fig:alltopologies}
\end{figure*}

The full-color SYM integrand can be written as
\begin{equation}
 {\cal A}_{5}^{(3)}=
-i g^9 \sum_{\sigma\in S_5}\sum_{i=1}^{42}
 \int\!\prod_{r=1}^{3} \frac{d^D\ell_r}{(2\pi)^D}\,
\frac{1}{S_i} \frac{C_iN_i}{\prod_{\alpha_i} P_{\alpha_i}^{\,2}}\, .
 \label{eq:gauge-integrand}
\end{equation}
Here $S_i$ is the automorphism factor and $C_i$ and $N_i$ are the color and
kinematic numerator factors. For each internal edge, the color identity
$C_s=C_t+C_u$ is accompanied by a dual Jacobi relation $N_s=N_t+N_u$ among the corresponding numerators, as illustrated in Fig.~\ref{fig:loopBCJ}. 
Using the no-triangle property of ${\cal N}=4$ SYM, we exclude all topologies containing triangle and bubble subgraphs. Under the full $S_5$ symmetry, there are 42 inequivalent topologies, shown in Fig.~\ref{fig:alltopologies}. 
We take the first two, (1) and (2), as the planar \emph{master topologies}. Imposing dual Jacobi relations on all four-point subgraphs then generates the remaining 40 numerators from the two masters.

To construct the ansatz for the master numerators, we expand them in the five-point supersymmetric prefactors $\beta_{ijklm}$ and $\gamma_{ij}$ introduced in \cite{Carrasco:2011mn}.
For the MHV superamplitude, the
definitions are
\begin{align}
\label{eq:beta12345}
& \beta_{12345} \equiv\delta^{(8)}(Q)\,
\frac{[12][23][34][45][51]}{4\,\varepsilon(1,2,3,4)}, \\
& \gamma_{12} = \beta_{12345} - \beta_{21345}
=\delta^{(8)}(Q)\,
\frac{[12]^{2}[34][45][35]}{4\,\varepsilon(1,2,3,4)} \,.\nonumber
\end{align}
Here $Q^{\alpha A}=\sum_{i} \lambda_i^\alpha \eta_i^{A}$ and
$\varepsilon(1,2,3,4)=\varepsilon_{\mu\nu\rho\sigma}
p_1^\mu p_2^\nu p_3^\rho p_4^\sigma$.
The $\gamma_{ij}$ factors obey $\sum_i \gamma_{ij}=0$ and $\gamma_{ij}=-\gamma_{ji}$, leaving six linearly
independent components.

The numerators are expanded in $\gamma_{ij}$ multiplied by Lorentz products of momenta: loop-loop contractions $(\ell_a \cdot \ell_b)$, loop-external contractions $(p_a \cdot \ell_b)$, and Mandelstam invariants $s_{ab}$.
The UV property of ${\cal N}=4$ SYM implies that a one-loop $n$-point subgraph carries at most $n-4$ powers of its loop momentum~\cite{Bern:2012uf}. For the two master numerators, 
this permits monomials of the form $s^2\gamma$,
$(p\cdot\ell)s\gamma$, $(p\cdot\ell)(p\cdot\ell)\gamma$, and $(\ell\cdot\ell)s\gamma$,
yielding 330 and 570 candidate terms, respectively.
Further momentum-weighted identities among the $\gamma_{ij} $~\cite{Carrasco:2011mn} 
must be quotiented out to avoid an overcomplete ansatz; 
an example quadratic in the Mandelstam variables is
\begin{equation}
\sum_{i=2}^4 s_{1i}^2 \gamma_{i5}
- (s_{13} s_{14} \gamma_{12} + s_{12} s_{14} \gamma_{13} + s_{12} s_{13} \gamma_{14}) = 0 .
\end{equation}
After quotienting, 281 and 491 independent monomials remain, respectively, giving 772 parameters in total.

\begin{figure}[t]
	\centerline{\includegraphics[height=5.2cm]{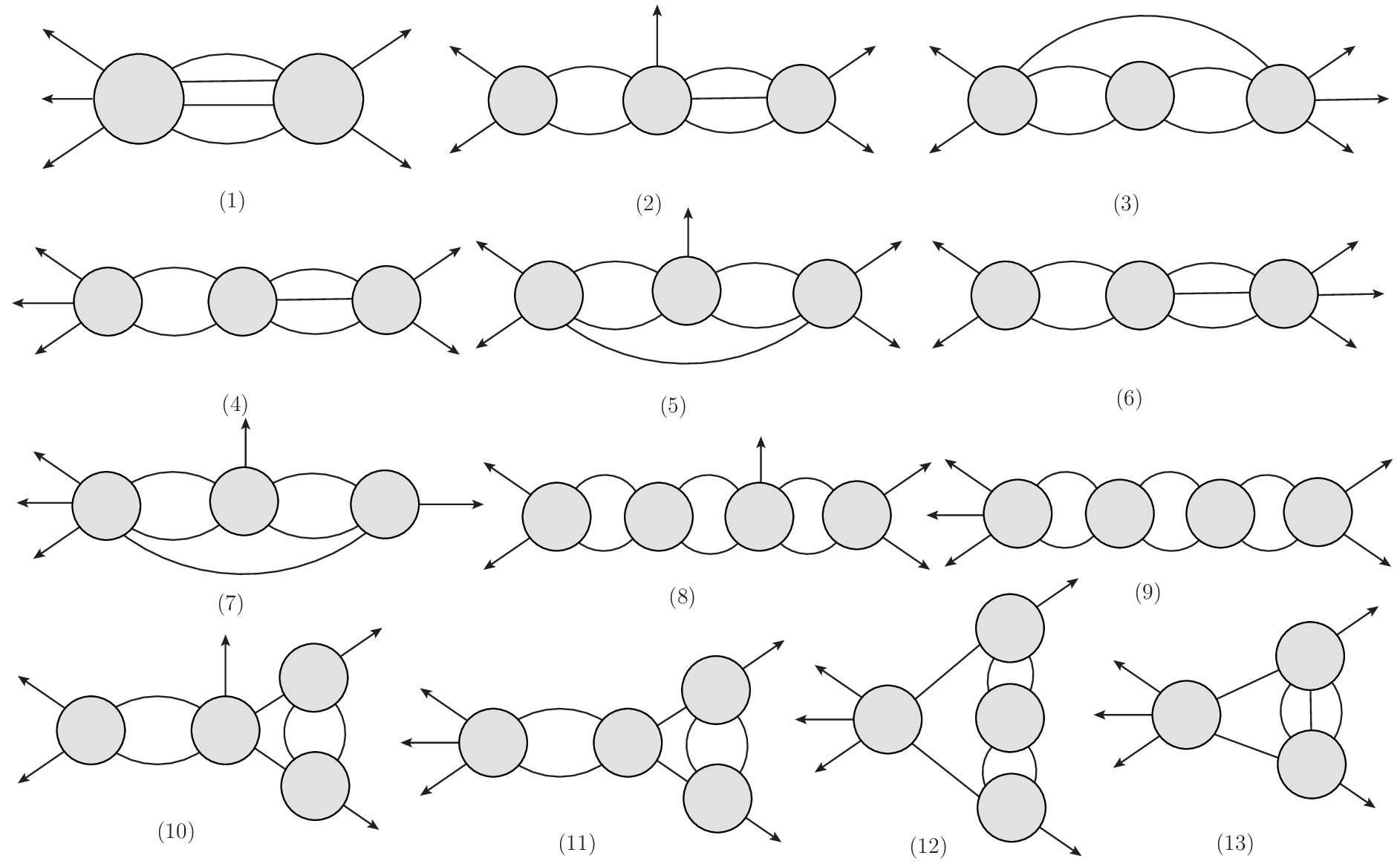} }
	\caption{The spanning set of 13 generalized-unitarity cuts  under the full $S_{5}$ symmetry.}
	\label{fig:cuts under S5}
\end{figure}

To fix these parameters, we first impose the $S_5$ automorphism symmetries of all topologies, reducing the 772 coefficients to 35.
We then impose the spanning set of 13 four-dimensional generalized-unitarity cuts, shown in Fig.~\ref{fig:cuts under S5}, which reduces the parameters to five.
Surprisingly, we find the first cut in Fig.~\ref{fig:cuts under S5} alone already produce this five-parameter family, and the remaining cuts, as well as all dual Jacobi relations, are then satisfied automatically.
The five free parameters cancel after integrand reduction using the method of \cite{Lin:2021qol}, confirming that they are generalized-gauge degrees of freedom.

\begin{figure}[t]
	\centerline{\includegraphics[height=1.5cm]{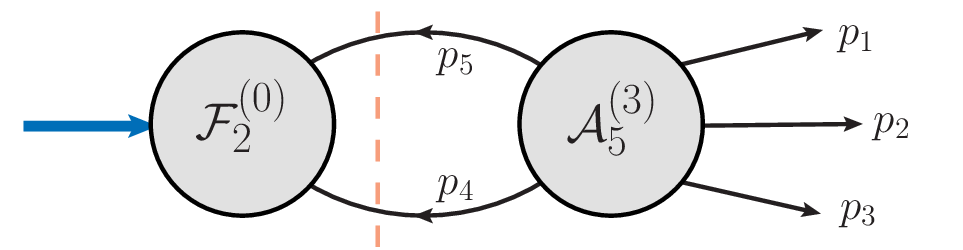} }
	\caption{The double cut for three-point form factor inspired the $S_3\times S_2$ construction.}
	\label{fig:double cut}
\end{figure}

We also construct the integrand independently using the smaller $S_3\times S_2$ symmetry motivated by the double-cut of the four-loop three-point form factor \cite{Lin:2021qol}, as shown in Fig.\,\ref{fig:double cut}.
In this construction, the external legs split into two sets $\{p_1,p_2,p_3\}$ and $\{p_4,p_5\}$. Permutations mixing the two sets generate additional distinct graphs and require a larger spanning set of unitarity cuts. The resulting solution contains 57 free parameters; 
after integrand reduction, it yields the same amplitude as the $S_5$-symmetric construction.

Although we use four-dimensional cuts, our result should capture the full $D$-dimensional loop-momentum information: (i) free parameters cancel under $D$-dimensional integrand reduction; (ii) the independent $S_5$ and $S_3\times S_2$ constructions yield the same result; (iii) the SYM UV pole agrees with the open-string prediction shown below. 
Further subtleties regarding $D$-dimensional external kinematics will be discussed later.

The three-loop five-point gravity amplitude follows directly via double copy:
\begin{equation}
\label{eq:M53loop-sum}
{\cal M}_5^{(3)} = \Big({\kappa\over2} \Big)^9 \sum_{S_5} \sum_{i=1}^{42} \int \prod_{r=1}^3 \frac{d^D\ell_r}{(2\pi)^D}\, \frac{1}{S_i} \,\frac{N_i \, \widetilde N_i}{\prod_{\alpha_i} P^2_{\alpha_i}} ,
\end{equation}
where $\widetilde N_i$ denotes the second gauge-theory copy. For
$\mathcal N=8$ SUGRA it is an identical $\mathcal N=4$ copy, 
with the two $\mathrm{SU}(4)$ R-symmetry index sets combining into $\mathrm{SU}(8)$. 
More generally, since only one copy needs to satisfy CK duality~\cite{Bern:2010ue}, we can pair $N_i$ with a valid cubic-graph representation of a less-supersymmetric gauge-theory integrand on the same graph basis, even if the $\widetilde N_i$ are not manifestly CK dual. This produces the corresponding gravity integrands with $\mathcal N<8$ supersymmetry.

For CK-dual solutions of both $S_5$ and $S_3\times S_2$ symmetries, the complete data for the master numerators, color factors, symmetry factors, propagators, 
and dual Jacobi relations are provided in the ancillary files.

\section{Ultraviolet structure}

\begin{figure}[t]
	\centerline{\includegraphics[height=2.cm]{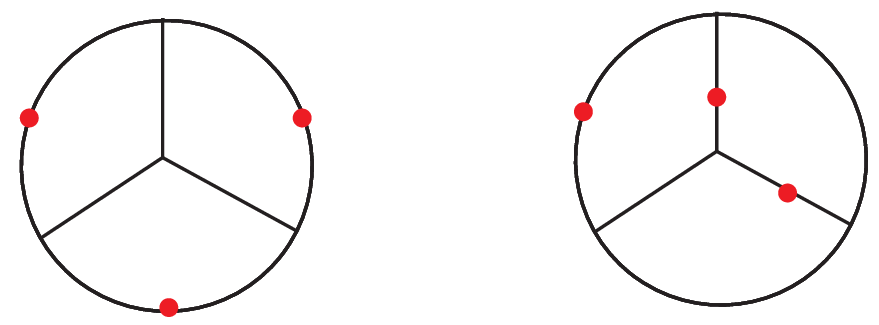} }
	\caption{The two three-loop vacuum integrals $V^{(A)}$ and $V^{(B)}$ that capture the UV divergence of the amplitudes. 
	Lines marked with red dots denote double propagators.}
	\label{fig:3loopvacuum}
\end{figure}

At three loops both amplitudes first diverge logarithmically at the critical dimension $D_c=6$,
consistent with the three-loop four-point result.
To extract the UV poles, we set all the external momenta to zero, reducing the amplitude to a sum of vacuum integrals; numerator tensors are reduced to scalar integrals via Lorentz invariance.
The resulting vacuum integrals are evaluated in $D=6-2\epsilon$, and their UV $1/\epsilon$ poles are isolated via IR rearrangement by assigning a fictitious mass to one propagator to regulate the infrared without modifying the UV counterterm \cite{Vladimirov:1979zm, Chetyrkin:1984xa}.
Only two scalar vacuum master integrals are needed, shown in Fig.~\ref{fig:3loopvacuum}. Their UV poles are  \cite{Bern:2008pv}
\begin{equation}
 V^{(A)}\big|_{\rm UV}=-\frac{1}{6(4\pi)^9\epsilon},\qquad
 V^{(B)}\big|_{\rm UV}=-\frac{\zeta_3-\tfrac13}{6(4\pi)^9\epsilon} .
 \label{eq:vacuum-integrals}
\end{equation}

\subsection{UV divergence in SYM}
\label{sec:SYMdiv}

Six of the 42 topologies contribute to the SYM UV divergence:
(37)--(42) in Fig.~\ref{fig:alltopologies},
which are precisely those containing two tree-level propagators. 
Their numerators are independent of the loop momenta, so they directly reduce to the vacuum integrals $V^{(A)}$ and $V^{(B)}$. 
Using the trace convention $\operatorname{Tr}_{1\ldots n}\equiv\operatorname{tr}(T^{a_{1}}\cdots T^{a_{n}})$, we obtain
\begin{align}\label{eq:A5l3_ct_SYM}
\mathcal{A}^{(3)}_5|_{\rm UV}
& =  - \frac{g^9 }{3 (4\pi)^9 \epsilon} \bigl( N_c^3 + 36 \zeta_3 N_c \bigr) \times \\
&  \Bigl[  \operatorname{Tr}_{12345}  \sum_{Z_5}  \Big(  \beta_{12345} + \frac{\gamma_{12}}{s_{12}} (s_{35} - 2 s_{12}) \Big)  + {\rm perms} \Bigr].\nonumber
\end{align}
The color dependence $(N_c^3 + 36 N_c \zeta_3)$ agrees with the three-loop four-point result~\cite{Bern:2008pv}.
Strikingly, all double-trace contributions cancel, although this is far from obvious at the integrand level.
The same cancellation was observed in the four-point amplitude \cite{Bern:2010tq} and is consistent with counterterm studies \cite{Berkovits:2009aw, Bossard:2009mn, Bjornsson:2010wm}.
The remaining single-trace kinematic structure coincides with that of the two-loop five-point divergence \cite{Carrasco:2011mn}.

Since $D_c >4$, the continuation to $D$ dimensions involves not only the loop momenta but also the \emph{external} states; the relevant amplitudes are those of the ten-dimensional theory reduced to $D$ dimensions. 
At four points, this continuation is direct since the external-state dependence factorizes into universal tree-level objects: $s t {\cal A}_4^{(0)}$ in SYM and $s t u {\cal M}_4^{(0)}$ in SUGRA \cite{BDDPR}.
At five points, however, no such overall tree-level factor exists and the lift is more subtle.
A candidate operation is the replacement $\gamma_{ij}\to\gamma_{ij}^{D}$, with $\gamma_{ij}^{D}$ the $D$-dimensional form introduced in \cite{Broedel:2011pd}:
\begin{align}
& \gamma^D_{12} =  \beta^D_{12345} - \beta^D_{21345} \,, \nonumber \\
& \beta_{12345}^{D} =
\frac{s_{12}s_{23}s_{34}s_{45}s_{51}}{16\,\mathcal{G}_{5}}
\Big[
2\,s_{14}s_{35}\, A^{(0)}_{(1,2,3,5,4)} 
\\
& \qquad\quad + (s_{15}s_{34}+s_{14}s_{35}-s_{13}s_{45})
A^{(0)}_{(1,2,3,4,5)} \Big] , \nonumber
\end{align}
where $A^{(0)}_{(i,j,k,l,m)}$ are color-ordered SYM tree amplitudes, and 
$\mathcal{G}_{5} \equiv \textrm{det}(p_i \cdot p_j)$, with $i, j=1,\ldots, 4$, is the five-point Gram determinant.
For five-gluon components, this replacement makes
Eq.~\eqref{eq:A5l3_ct_SYM} agree prefectly with the result predicted by the open-string expansion at generic kinematics \cite{Schlotterer:2012rc, Mafra:2015mja}.
However, as we will see below, this success does not proceed to the gravity case.

\subsection{UV divergence in SUGRA}
\label{sec:SUGRAdiv}

For gravity, after the double copy, more topologies contribute to the UV poles, 
and their numerators require tensor reduction through rank four. The resulting pole is 
\begin{align}
\label{eq:A5l3_ct_SUGRA}
& \mathcal{M}^{(3)}_5|_{\rm UV}
= - i \Big({\kappa\over2} \Big)^9 \frac{5 \zeta_3}{12 (4\pi)^9 \epsilon} \sum_{S_5}  {\gamma}_{12}\widetilde{\gamma}_{12} \\
&
\times \bigg(
{s_{34}s_{35}s_{45} \over s_{12} }  +\frac{3}{8} \Big[s_{12}^2+s_{34}^2+s_{35}^2+s_{45}^2
\nonumber \\
&
\qquad -(s_{13}-s_{23})^2
-(s_{14}-s_{24})^2
-(s_{15}-s_{25})^2
\Big] \bigg)   , \nonumber
\end{align}
where $\widetilde{\gamma}_{ij} = \gamma_{ij}|_{\eta_i^A \rightarrow \eta_i^{A+4}}$.
For four-dimensional kinematics, Eq.~\eqref{eq:A5l3_ct_SUGRA} agrees with the closed-string-inspired expression \cite{Schlotterer:2012rc}:
\begin{align}\label{eq:A5l3_ct_SUGRA_string}
\mathcal{M}^{(3)}_5|^{\rm string}_{\rm UV}
&= -i \Big({\kappa\over2} \Big)^9 \frac{20 \zeta_3}{(4\pi)^9 \epsilon}\\
&
\times \begin{pmatrix}
A^{(0)}_{(1,2,3,5,4)}\\[2mm]
A^{(0)}_{(1,3,2,5,4)}
\end{pmatrix}^T
\cdot S_0 \cdot
(M_{3})^2 \cdot
\begin{pmatrix}
A^{(0)}_{(1,2,3,4,5)}\\[2mm]
A^{(0)}_{(1,3,2,4,5)}
\end{pmatrix},\nonumber
\end{align}
where $S_0$ is the KLT kernel and $M_3$ is a matrix of Mandelstam invariants \cite{Schlotterer:2012rc}.

However, when the replacement $\gamma_{ij}\to\gamma_{ij}^{D}$ is evaluated with generic $D$-dimensional kinematics, 
a difference appears:
\begin{equation}\label{eq:M53loop_evanescentdiff}
\begin{aligned}
&\mathcal{M}_5^{(3)}\bigr|^{\gamma^D}_{\rm UV} - \mathcal{M}^{(3)}_5|^{\rm string}_{\rm UV} =\\
&- i \Big({\kappa\over2} \Big)^9 \frac{5 \zeta_3}{48 (4\pi)^9 \epsilon} {s_{12}s_{13}s_{15}s_{23}s_{24}s_{34}s_{45} \over \mathcal{G}_{5}} \Big(\sum_{i<j}s_{ij}^3\Big) {\cal E}
\end{aligned}
\end{equation}
where ${\cal E}$ is the \emph{evanescent} combination:
\begin{align}
{\cal E} = & \big(A^{(0)}_{(1,2,3,4,5)}\big)^2 s_{12}s_{34} + \big( A^{(0)}_{(1,3,2,4,5)} \big)^2 s_{13}s_{24} \\
+
& A^{(0)}_{(1,2,3,4,5)} A^{(0)}_{(1,3,2,4,5)} (s_{12}s_{34} + s_{13}s_{24} - s_{14}s_{23}) \,. \nonumber
\end{align}
This quadratic combination of amplitudes vanishes in four dimensions but is nonzero for generic $D$-dimensional states.
The same ${\cal E}$ appears in the analogous one- and two-loop comparisons, dressed by
Mandelstam polynomials of successively lower degree: for the one- and the two-loop cases \cite{Carrasco:2011mn}, one simply replaces $(\sum s_{ij}^3)$ in \eqref{eq:M53loop_evanescentdiff} by a constant and $(\sum s_{ij}^2)$, respectively.
This suggests that the discrepancy may have a systematic simple pattern for general loops.

\section{Summary and discussion}

We have constructed the complete three-loop five-point integrand in $\mathcal{N}=4$ SYM in a manifestly CK-dual form,
and its double copy directly yields the corresponding $\mathcal{N}=8$ amplitude.
We have also computed the UV divergences of both amplitudes at their critical dimension $D_c=6$.
Several features of these results deserve comment.

First, the five-point SYM UV pole probes the full non-linear completion of the linearized $\partial^2 F^4$ counterterm operator, of the schematic form $D^2F^4 + F^5$.
Remarkably, we find that the simple two-term operator 
\begin{equation}
{\cal O}_{\rm ct} = {\rm tr}(D_\mu F_{\nu \rho} D_\sigma F^{\nu}_{~\lambda} F^{\lambda\mu} F^{\rho\sigma})
- {\rm tr}(D_\mu F_{\nu \rho} D_\sigma F^{\nu}_{~\lambda} F^{\rho\sigma} F^{\lambda\mu} )
\end{equation}
reproduces the UV divergences of both four- and five-point amplitudes.
At on-shell level, ${\cal O}_{\rm ct}$ is equivalent to the superstring effective action at $\alpha'^3$ order, e.g.,~the eight-term form first obtained in \cite{Koerber:2001uu} (see also \cite{Barreiro:2012aw}).

Second, our results sharpen the question of how four-dimensional integrand representations are lifted to $D$ dimensions.
The replacement $\gamma_{ij}\to\gamma^D_{ij}$ is a natural operation 
that packages the external-state dependence into $D$-dimensional tree amplitudes, 
and it reproduces the SYM pole. 
However, the SUGRA mismatch suggests that it does not determine the complete $D$-dimensional integrand.  
A natural explanation is the absence of parity-odd contractions involving $\epsilon_{\mu_1\ldots\mu_D}$ with $D>4$, 
which do not enter our CK-dual construction. 
Further insight into the lifting issue, as well as the origin of the inverse Gram determinant, comes from comparing with the two-loop five-point pure-spinor construction \cite{Mafra:2015mja}, whose ten-dimensional numerators are fully local and contain tensorial building blocks such as $T^m_{i,j,k|l,m}$ with $m$ a ten-dimensional Lorentz index.  
Projecting a local
vector $T^m$ onto four independent external momenta gives
\begin{equation}
 T^m=p_i^m(G^{-1})_{ij}(p_j\!\cdot T)+T_\perp^m,
 \qquad p_i\!\cdot T_\perp=0,
 \label{eq:tensor-projection}
\end{equation}
where $G_{ij}=p_i\cdot p_j$ with $i,j=1,\ldots,4$. 
The inverse Gram matrices arise from solving for projected coefficients,
while $T_\perp^m$ is the component transverse to the four dimensions. 
Such transverse pieces do not contribute to the SYM UV pole in generic $D$ dimensions.
In gravity, however, the product $N_i\widetilde N_i$ allows cross terms and transverse
bilinears to survive.

As for outlook, a three-loop extension of the two-loop pure-spinor construction \cite{Mafra:2015mja} would clearly be valuable.
It would also be interesting to explore double copies with reduced supersymmetry based on the numerators obtained here.
Our results provide the complete non-planar integrand data needed for integration (see \cite{Chicherin:2025mvc} for recent progress on planar integrals).
Finally, it would be interesting to extend the construction to five points at four loops and beyond.

\vskip .3cm

{\it Acknowledgments.}
This work is supported by the National Natural Science Foundation of
China (Grants No.~12425504, 12447101, 12247103) and by the Chinese
Academy of Sciences (Grant No.~YSBR-101).
We also acknowledge the support of the HPC Cluster of ITP-CAS.

\bibliographystyle{apsrev4-2}

\bibliography{ff}

\end{document}